\documentclass[prl,reprint,nobibnotes, aps,showpacs]{revtex4-1}

\usepackage[english]{babel}
\usepackage{grffile}
\usepackage{graphicx,epstopdf,color}
\usepackage{amsmath,amssymb,psfrag}
\usepackage[caption=false]{subfig}

\usepackage[draft]{fixme}
\usepackage[latin1]{inputenc}

\newcommand{\pcal}{\mathcal{P}}
\newcommand{\vtss}{v^{\text{TSS}}}
\newcommand{\peq}{\mathcal{P}^{\text{eq}}}
\newcommand{\Pout}{P_{\text{out}}}
\newcommand{\D}[1]{\partial_{#1}}
\newcommand{\FE}{\mathcal{F}}
\newcommand{\dup}{\text{d}}

\begin{document}

\title{Entropy-generated power and its efficiency}
\author{Natalia Golubeva}
\author{Alberto Imparato}
\affiliation{Department of Physics and Astronomy, University of Aarhus, DK--8000 Aarhus C, Denmark}
\author{Massimiliano Esposito}
\affiliation{Complex Systems and Statistical Mechanics, University of Luxembourg, L--1511 Luxembourg, Luxembourg}
\date{\today}

\begin{abstract}
We propose a simple model for a motor that generates mechanical motion by exploiting an entropic force arising from the topology of the underlying phase space. We show that the generation of mechanical forces in our system is surprisingly robust to local changes in kinetic and topological parameters. Furthermore, we find that the efficiency at maximum power may show discontinuities. 
\end{abstract}   

\maketitle
 
The framework of equilibrium thermodynamics provides several examples of entropic forces including osmotic pressure, the hydrophobic force in aqueous solutions and the elastic forces arising in models of freely-jointed polymers such as rubber \cite{Nelson2007}. More recently, physicists have started studying entropic effects in out-of-equilibrium phenomena such as transport of particles through channels \cite{Reguera2006}, and exploited the entropic forces to construct an efficient sorting mechanism \cite{Reguera2012}. These processes require energy from an external driving force. Entropic effects in copolymerization mechanisms have also been considered \cite{Andrieux2008a}. There, the thermodynamic affinity driving these nonequilibrium processes results from the interplay between the information stored in the monomer sequence and the free energy per monomer. This affinity contains an entropic contribution which depends on the enthalpic force and is thus not accessible to direct experimental manipulation \cite{Esposito2010a}. 

\begin{figure}
\centering
\includegraphics[width=.7\columnwidth]{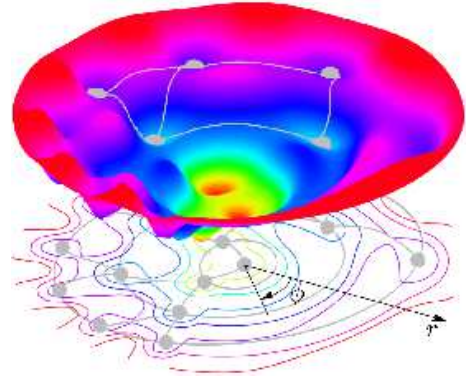}
\caption{A generic potential energy landscape in two dimensions with exponentially increasing accessible phase space associated to the radial direction $r$. The motion along $r$ is subject to an average negative force hindering the motion. Diffusion on the potential energy surface can be described by a Markovian jumping process on the corresponding two-dimensional network.}
   \label{fig:model}
 \end{figure}

In this Letter we propose to use entropy to fuel an engine that generates mechanical motion in an out-of-equilibrium setting. For this purpose we introduce a simple model in which the thermodynamic driving force is of a purely entropic nature and directly controllable. We consider a Markovian jumping process on a two-dimensional network as a model of diffusion in a generic landscape like the one depicted in fig. \ref{fig:model} with exponentially increasing available phase space along the direction of positive average slope. We show analytically using stochastic thermodynamics and coarse-graining that, with few restrictions on the topology and kinetics of the process, the system exhibits a positive steady-state velocity in the radial direction when a time-scale separation between the two directions exists. The resulting dynamics can thus be interpreted as that of a motor pumping particles uphill against a mechanical gradient powered by a purely entropic force originating from the increasing phase space volume of the slow coordinate. Furthermore, we show numerically that the pumping effect is surprisingly robust even when the time scale separation assumption is not strictly fulfilled. Finally, we study the efficiency at maximum power (EMP) of our model motor and show that the qualitative behaviour of this quantity can differ severely from the one observed in one-dimensional, isothermal engine models traditionally considered in the context of EMP \cite{Golubeva2012,VandenBroeck2012,Seifert2011a}. Especially, we find that the EMP can exhibit discontinuities.
   
\paragraph{Model.} 
We consider a diffusion process on a two-dimensional potential energy landscape as depicted in fig. \ref{fig:model}. The continuous phase space model can be approximately mapped into a Markovian jumping process on a network with the local minima of the potential energy landscape defining the nodes or microstates of the network, and the minimal energy paths between the minima corresponding to the edges or reversible transitions between microstates, as illustrated schematically in fig. \ref{fig:model} \cite{Gfeller2007}. For simplicity, we therefore restrict the discussion to a discrete Markov process on the resulting two-dimensional network model. However, our analysis can easily be extended to systems with continuous phase space such as the billiard model studied in \cite{Barra2006}, and all the conclusions of the Letter thus apply to this class of systems as discussed in the Appendix.

We denote the microstates of the radially symmetric network by $m=(r,\phi)$, where the slow radial coordinate $r$ can be thought of as a reaction coordinate that represents the macroscopic level of description after coarse-graining. The number $N_r$ of microstates comprising the available phase space of a given macrostate $r$ increases exponentially as $N_r=k^r$, where $k$ is some integer. The transition rate for going from state $m$ to state $m'$ is denoted by $w_{mm'}$. For simplicity, we consider a potential that is constant along the $\phi$-direction and increases linearly in the radial direction. In this case, the local detailed balance condition for transitions between macrostates $r$ and $r'=r+1$ takes the form  $w_{mm'}/w_{m'm}=\exp(-\beta\Delta E)$, where $\beta=1/T$ is the inverse temperature, and the mechanical gradient $\Delta E>0$ is independent of $r$. The Boltzmann constant $k_B$ is set to unity. 

The evolution of the probability $P_r$ to find the system in macrostate $r$ is governed by the coarse-grained master equation \cite{Esposito2012b} 
\begin{equation*}
  \label{eq:mastereq}
  \dot{P}_r=-(W_{r,r+1}+W_{r,r-1})P_r+W_{r+1,r}P_{r+1}+W_{r-1,r}P_{r-1}.
\end{equation*}
Here, the effective transition rates $W_{r,r\pm 1}$ for going from $r$ to $r\pm 1$ are obtained by summing over the relevant transitions between microstates, i.e.
\begin{equation}
  \label{eq:weff}
  W_{r,r\pm 1}=\sum_{\phi_r,\phi_{r'}, r'=r\pm 1} w_{(r,\phi_r)(r',\phi_{r'})} \pcal_{\phi_r|r},
\end{equation}
where $\pcal_{\phi_r|r}$ is the conditional probability to find the system in microstate $(r,\phi_r)$ given that the macrostate is known to be $r$. For fast dynamics along the $\phi$-coordinate, this probability attains the equilibrium value $\pcal^{\text{eq}}_{\phi_r|r}=1/N_r=k^{-r}$, and eq. \eqref{eq:weff} becomes
\begin{equation}
  W_{r,r\pm 1}=k^{-r} n_{r,r\pm 1}\langle w_{mm'} \rangle_{n_{r,r\pm 1}},
\label{eq:wcoarse}
\end{equation}
where $n_{r,r'}$ is the number of connections between state $r$ and $r'$, and $\langle w_{mm'} \rangle_{n_{r,r'}}$ denotes the average rate for these transitions. The modified local detailed balance relation after coarse-graining thus reads
\begin{equation}
  \label{eq:wratio}
  \frac{W_{r,r+1}}{W_{r+1,r}}=k \frac{\langle w_{mm'} \rangle_{n_{r,r+1}}}{\langle w_{mm'} \rangle_{n_{r+1,r}}}=e^{-\beta \Delta F},
\end{equation}
where we have introduced the change in the free energy $F$,
\begin{equation}
  \label{eq:wratio2}
  \Delta F=\Delta E-T\Delta S=\Delta E-T\log(k),
\end{equation}
associated with a transition from $r$ to $r+1$. From eqs. \eqref{eq:wratio}--\eqref{eq:wratio2} it is evident that the entropic force $T\Delta S=T\log(N_{r+1}/N_r)=T\log(k)$ arising from the topology of the network is able to generate motion against an external bias $\Delta E<T\log(k)$ whenever the available coarse-grained phase space increases exponentially, i.e. $N_r=k^r$. However, the critical value $k^*$ required to generate motion increases exponentially with the opposing force as $k^*=\exp[\Delta E/T]$. We emphasize that our formalism is also valid in the more general case of a position dependent force $\Delta E(r)$. In this case, the rate of the exponential increase of $N_r$ has to be chosen such that $\Delta F(r)>0$ for all $r$.
It is also interesting to note that the coarse-graining of the purely energetic microscopic network dynamics leads to a description in terms of nonequilibrium free energies at the macroscopic level as discussed in \cite{Esposito2012b}.

\begin{figure}
\centering
\includegraphics[width=.7\columnwidth]{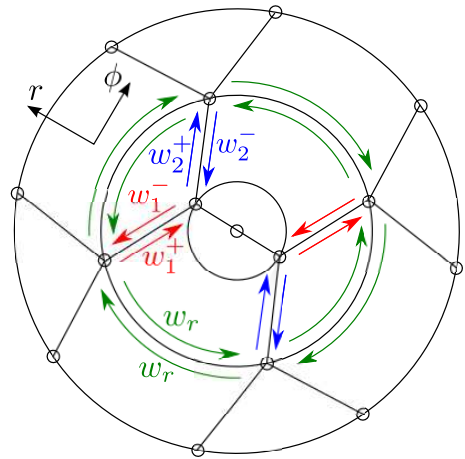}
\caption{In the model network each state in layer $r$ is connected to two new states in $r+1$ by two edges with transition rates $w_1^\pm$ and $w_2^\pm$, respectively. Transitions between neighbouring states in $r$ take place with rates $w_r$.}
   \label{fig:modelasymm}
 \end{figure}

If, moreover, each macrostate $r$ is connected to the state $r+1$ by $k^{r+1}$ transitions, i.e., $n_{r,r+1}=k^{r+1}$, such that every microstate in $r$ is on average connected to $k$ microstates in $r+1$, see e.g. fig. \ref{fig:modelasymm}, and the average transition rate $\langle w_{mm'} \rangle_{n_{r,r'}}$ is independent of $r$, the velocity becomes position independent in the long-time limit, and the system reaches a nonequilibrium steady-state. In this case, the two-dimensional problem with fast equilibration along the $\phi$-direction reduces to a one-dimensional asymmetric random walk along the $r$-direction with the steady-state velocity given by 
\begin{equation}
\vtss_r=W_{r,r+1}-W_{r+1,r}=\langle w_{mm'} \rangle_{n_{r,r-1}} (e^{-\beta \Delta F}-1),
\label{eq:v}
\end{equation}
as follows from eqs.~(\ref{eq:wcoarse})-(\ref{eq:wratio}). In the following we discuss the conditions under which the strict mathematical time scale separation (TSS) leading to eqs. \eqref{eq:wcoarse}--\eqref{eq:v} is valid and show that steady-state operation can be obtained even when the assumption of TSS is relaxed.

\begin{figure}
  \psfrag{xt}[ct][ct][1.]{$\langle r \rangle$}
  \psfrag{t}[ct][ct][1.]{$t$}
  \psfrag{legendlegend1}[rr][rr][1.]{$\Delta E=0.1 \Delta S$, $n=20$}
  \psfrag{legend1}[rr][rr][1.]{$n=21$}
  \psfrag{legendlegend2}[rr][rr][1.]{$\Delta E=0.5 \Delta S$, $n=20$}
  \psfrag{legend2}[rr][rr][1.]{$n=21$}
  \psfrag{legendlegend3}[rr][rr][1.]{$\Delta E=1.5 \Delta S$, $n=20$}
  \psfrag{legend3}[rr][rr][1.]{$n=21$}
  \psfrag{legendfit}[rr][rr][1.]{fit, eq. \eqref{eq:v2}}
 \centering
 \includegraphics[width=.9\columnwidth]{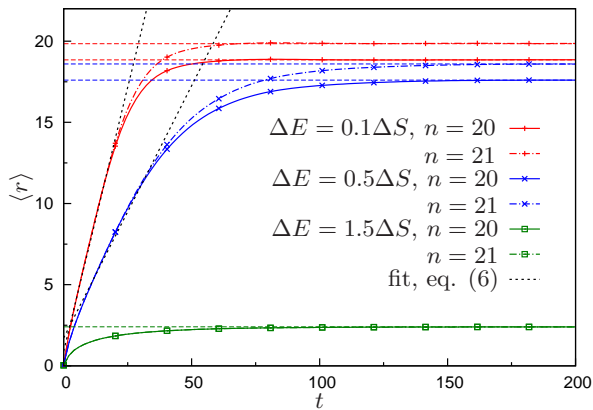}
 \caption{The average position $\langle r \rangle$ along the $r$-coordinate as a function of time for various values of the opposing force $\Delta E$ and two different system sizes $n$ as specified in the legend. The dashed (black) lines correspond to the analytical expression, eq. \eqref{eq:v2}, for the steady-state velocity. For each curve, the horizonal dashed line denotes the equilibrium value of $\langle r\rangle$. Parameter values are $w_1^-=1$, $w_2^-=0.5$, $w_{1,2}^+=w_{1,2}^-e^{-\Delta E}$ and $w_r=2^{2r}$. The values of $n$ are chosen such that the simulation times lie within a reasonable range.}
    \label{fig:xttss}
  \end{figure}

\paragraph{Requirement of Time Scale Separation.}
To explore the relation of time scales it is convenient to consider a model system as depicted in fig. \ref{fig:modelasymm}. Each state in $r$ is connected to exactly two states in $r+1$, i.e. $k=2$, and we denote by $w_i^+$ ($w_i^-$), with $i=1,2$, the microscopic forward (backward) transition rate which is indedendent of $r$. The dynamics within a given layer $r$ corresponds to a symmetric random walk on a finite lattice with $N_r=2^r$ sites and jumping rates equal to $w_r$. The slowest eigenvalue $\lambda_r$ governing the diffusive relaxation towards equilibrium within $r$ thus scales as $\lambda_r \sim w_r N_r^{-2}=w_r 2^{-2r}$. Consequently, in order to ensure strict TSS between the dynamics along the $r$ and $\phi$-directions, the rates along $\phi$ must scale as $w_r=w_1 2^{2r}$ with the first-layer rate fulfilling $w_1 \gg \max_{i=1,2}(w_i^++w_i^-)$. Under these assumptions the velocity in the long-time limit is given by eq. \eqref{eq:v},
\begin{equation}
  \vtss_r=\frac{1}{2}(w_1^-+w_2^-) (e^{-\beta (\Delta E-T\Delta S)}-1),
  \label{eq:v2}
\end{equation}
where $\Delta S=\log(2)$. The prediction of eq. \eqref{eq:v2} is verified by numerical simulations as shown in fig. \ref{fig:xttss}, where we plot the average position $\langle r \rangle $ as a function of time for different values of the slope $\Delta E$. For $\Delta E<T\Delta S$, the dynamics exhibits a short transient regime followed by a linear increase in position with the slope given by eq. \eqref{eq:v2}, as expected. Finally, the velocity decreases due to finite-size effects as $\langle r \rangle$ approaches its equilibrium value. The extent of the linear regime increases with the system size as illustrated in the figure by considering two different values of the system radius $n$. For $\Delta E>T\Delta S$, however, no steady-state regime is present, and the equilibrium distribution as well as the approach to equilibrium is independent of the system size. 

 \begin{figure}
  \psfrag{xt}[ct][ct][1.]{$\langle r \rangle$}
  \psfrag{t}[ct][ct][1.]{$t$}
  \psfrag{leg1}[rr][rr][1.]{$f=0.9$, $w_r=2^{2r}$}
  \psfrag{leg2}[rr][rr][1.]{$w_r=1$}
  \psfrag{leg3}[rr][rr][1.]{$w_r=0$}
  \psfrag{leg4}[rr][rr][1.]{$f=0.1$, $w_r=2^{2r}$}
  \psfrag{leg5}[rr][rr][1.]{$w_r=1$}
  \psfrag{leg6}[rr][rr][1.]{$w_r=0$}
  \psfrag{leg7}[rr][rr][1.]{$f=0.001$, $w_r=2^{2r}$}
  \psfrag{leg8}[rr][rr][1.]{$w_r=1$}
  \psfrag{leg9}[rr][rr][1.]{$w_r=0$}
 \centering
 \includegraphics[width=.9\columnwidth]{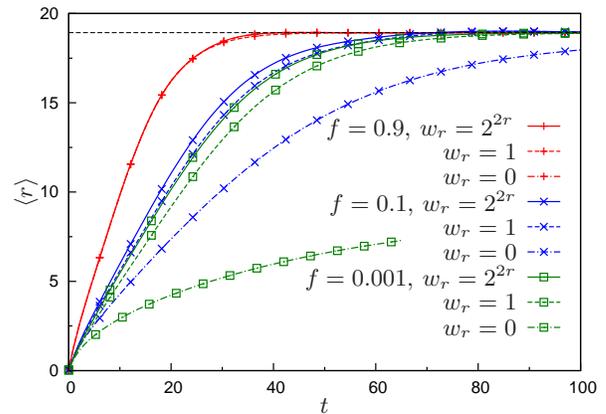}
 \caption{The average position $\langle r \rangle$ along the $r$-coordinate as a function of time for various values of the asymmetry parameter $f=w_2^\pm/w_1^\pm$ and for various time scales along the $\phi$-direction quantified by $w_r$ as specified in the legend. The dashed line denotes the equilibrium value of $\langle r\rangle$ for $\Delta E=0.05 \Delta S$ and $n=20$. Parameters are $w_1^-=1$ and $w_1^+=e^{-\Delta E}$.}
    \label{fig:xtnotss}
  \end{figure}
In order to explore the robustness of the entropic motor we performed numerical simulations of the system for different values of the rates $w_r$ that do no strictly satisfy the TSS condition and for different system asymmetries quantified by the ratio $f=w_2^\pm/w_1^\pm<1$. In fig. \ref{fig:xtnotss} we plot the time evolution of $\langle r \rangle$ when all the rates $w_r$ are of the same order of magnitude as the fastest transition rates $w_1^\pm$ along $r$ ($w_r=1$, dashed lines). For comparison we also show the resulting dynamics in the case of complete TSS (solid lines) and in the absence of connections along $\phi$ ($w_r=0$, dot-dashed lines). We note that in the special case $w_1^\pm=w_2^\pm$ ($f=1$), the equilibration along the $\phi$-direction is fulfilled automatically due to the symmetry of the problem, and the velocity is thus independent of $w_r$. For small asymmetries we therefore expect $\langle r \rangle$ to depend weekly on $w_r$ as shown in fig. \ref{fig:xtnotss} ($f=0.9$, plus symbols). In more asymmetric systems the entropic force is clearly unable to generate steady-state motion for vanishing $w_r$. However, strikingly, even for very high asymmetries ($f=0.001$, square symbols), the merely moderately strong connections $w_r=1$ ensure an approximately constant velocity that is close to the optimal value $\vtss$, eq. \eqref{eq:v2}. Such a robustness is attributed to the interplay of local topology and the fast decaying eigenmodes of the dynamics along $\phi$. The density fluctuations imposed by the asymmetric kinetic rates exhibit a periodic structure due to the self-similar geometry of the problem and thus decay on time scales shorter than the typical times $1/w_{1,2}^\pm$ for movement between macrostates. As a result, the conditional probability distribution, $\pcal_{\phi_r|r}$, along $\phi$ becomes quasi-stationary with an associated entropy $S'\lesssim S=\log(N_r)=r\log(k)$, which in turns yields a nearly constant velocity $v' \lesssim \vtss$ obtained by replacing $\Delta S$ with $\Delta S'$ in eq. \eqref{eq:v2}. Based on these considerations we therefore conclude that a steady power output can be generated by entropic forces for a wide range of network topologies and transition rate distributions as long as these do not systematically build up density inhomogenuities in $\pcal_{\phi_r|r}$ that decay slower than the characteristic time scales for microscopic transitions along the slow coordinate.  

\paragraph{Efficiency at maximum power.}
We now turn to the question of how efficiently the proposed entropic motor is performing mechanical work. The system is by construction tightly coupled \cite{Golubeva2012,Seifert2011a,Golubeva2012a,Qian2004}, and the efficiency $\eta=v \Delta E/(v T\Delta S)$ of the motor is thus independent of its velocity $v$. The efficiency attains the maximum value 1 at thermodynamic equilibrium corresponding to a vanishing free energy, $\Delta F=0$, and velocity, $v=0$. In this quasistatic limit the power output $\Pout=v \Delta E$ is thus zero and therefore of limited practical interest. We hence turn to calculating the efficiency at maximum power (EMP), which is obtained by optimizing the output power with respect to the opposing force $\Delta E$ with the velocity given by eq. \eqref{eq:v2}. In order to fully specify the problem we introduce the Arrhenius parametrization of the jumping rates, $w_i^\pm=w_0\exp[\beta(-\Delta E_i^0 \mp  \Delta E \theta_i^\pm)]$ for $i=1,2$, which is appropriate for describing diffusion over a high barrier \cite{vanKampen}. Here, $\Delta E_i^0$ is the barrier height for pathway $i$ in the absence of force, i.e. for $\Delta E=0$, and $w_0$ is a microscopic rate constant assumed to be the same for both transitions. The so-called load sharing factors $\theta_i^\pm$ denote the position of the energy barrier or transition state for motion in the forward and backward directions, respectively, and must fulfill $\theta_i^++\theta_i^-=1$ in order to satisfy local detailed balance. 
 \begin{figure}
 \psfrag{l1}[rr][rr][1.]{{\scriptsize $1$}}
 \psfrag{l2}[rb][rb][1.]{{\scriptsize $2$}}
 \psfrag{l3}[rb][rb][1.]{{\scriptsize $3$}}
 \psfrag{l4}[rb][rb][1.]{{\scriptsize $4$}}
 \psfrag{l5}[rb][rb][1.]{{\scriptsize $5$}}
 \psfrag{l6}[rb][rb][1.]{{\scriptsize $6$}}
 \psfrag{l7}[rb][rb][1.]{{\scriptsize $7$}}
 \psfrag{emp}[ct][ct][1.]{$\eta^*$}
 \psfrag{pout}[ct][ct][1.]{$\Pout$}
 \psfrag{DS}[ct][ct][1.]{$\Delta S$}
 \psfrag{DE}[ct][ct][1.]{$\Delta E$}
 \psfrag{DS1}[lt][lt][1.]{{\scriptsize $\Delta S=5$}}
 \psfrag{DS2}[ct][ct][1.]{{\scriptsize $\Delta S=11$}}
 \psfrag{DS3}[lt][lt][1.]{{\scriptsize $\Delta S=13$}}
 \psfrag{a}[cB][cB][1.]{{\scriptsize $\textbf{(a)}$}}
 \psfrag{b}[cB][cB][1.]{{\scriptsize $\textbf{(b)}$}}
 \psfrag{c}[cB][cB][1.]{{\scriptsize $\textbf{(c)}$}}
 \psfrag{d}[cB][cB][1.]{{\scriptsize $\textbf{(d)}$}}
 \psfrag{blabl1}[cc][cc][1.]{{\footnotesize $w=\infty$}}
 \psfrag{blabl2}[cc][cc][1.]{{\footnotesize $w=0$}}
 \psfrag{blabl3}[cc][cc][1.]{{\footnotesize $w=2$}}
 \psfrag{blabl4}[cc][cc][1.]{{\footnotesize $w=4$}}
 \psfrag{blabl5}[cc][cc][1.]{{\footnotesize $w=4.4$}}
 \psfrag{blabl6}[cc][cc][1.]{{\footnotesize $w=4.6$}}
 \psfrag{blabl7}[cc][cc][1.]{{\footnotesize $w=5$}}
 \psfrag{blabl8}[cc][cc][1.]{{\footnotesize $w=8$}}
 \centering
 \includegraphics[width=\columnwidth]{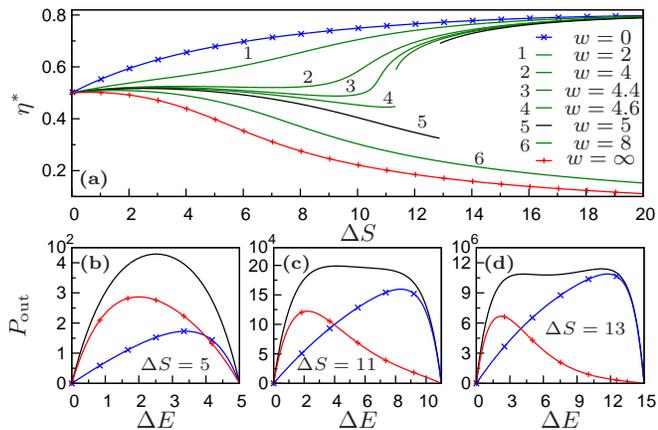}
 \caption{(a) The efficiency at maximum power $\eta^*$ for entropy-generated power as a function of the entropic driving $\Delta S$ for different barrier height ratios parametrized by $w=\exp[\beta(\Delta E_1^0-\Delta E_2^0)]$. (b)-(d) Power output $\Pout$ (green, solid line) as a function of the load $\Delta E$ for different values of the input $\Delta S$ for $w=5$. The crosses (blue) and the plus symbols (red) denote the contributions from the first and second pathway, respectively. For small $\Delta S$, the maximum of $\Pout$ is dictated by the second pathway (b)-(c). As $\Delta S$ increases, the first pathway becomes dominant (d). The values of the load factors are $\theta_1^+=0.045$ and $\theta_2^+=0.45$ respectively.}
\label{fig:emp}
\end{figure}

For a given driving $\Delta S$, the EMP $\eta^*$ is calculated as $\eta^*=\Delta E^*/(T\Delta S)$, where the optimal load force $\Delta E^*$ fulfills $\text{d}\Pout/\text{d}\Delta E=0$. In the linear regime, $\Delta S \ll 1$, the EMP to first order in $\Delta S$ is given by
\begin{equation}
  \label{eq:lr}
  \eta^*=\frac{1}{2}+\left[\left(\frac{1}{2}-\theta_1^+ \right)+w\left( \frac{1}{2}-\theta_2^+ \right) \right] \frac{\Delta S}{8(1+w)},
\end{equation}
where $w=\exp[\beta(\Delta E_1^0-\Delta E_2^0)]$ quantifies the difference in the zero-force barriers for the pathways. Especially, our model yields the linear response value $1/2$ for the EMP, which is universal for systems with well-defined thermodynamic forces as discussed in \cite{Golubeva2012,Esposito2009,VandenBroeck2012,Seifert2011a}. Hence, the entropic force in our system is a true thermodynamic force, while in the model of copolymerization \cite{Andrieux2008a} the entropic driving force depends on the extracted work and is thus not the appropriate quantity to fix when calculating EMP \cite{Esposito2010a}. For $w=0,\infty$ the first-order term in eq. \eqref{eq:lr} reduces to the well-known result for one barrier \cite{Golubeva2012,Esposito2009,VandenBroeck2012,Seifert2011a}.

Far from equilibrium, however, the EMP exhibits very different behaviour depending on the choice of parameters. Specifically, if the physical nature of the two pathways gives rise to significantly different activation barrier positions, i.e. the load parameters satisfy $\theta_1^+ \ll \theta_2^+$, the EMP exhibits a discontinuity as a function of $\Delta S$ for a range of values of $w$ as illustrated in fig. \ref{fig:emp}a \footnote{We note that the condition $\theta_1^+ \ll \theta_2^+$ can only be fulfilled, if process 1 is Eyring-like. Process 2, however, should not necessarily satisfy the Kramers-like limit \cite{VandenBroeck2012}.}. For $w \lesssim 1$, i.e. even for comparable zero-force barrier heights, the EMP $\eta^*$ is continuous and largely equivalent to the one obtained for one barrier with load parameter $\theta_1^+$ (crosses), since for small $\theta^+$ the barrier lies close to the initial state and is thus little affected by the external load. For larger values of $w$, however, the second barrier located at $\theta_2^+$ becomes low enough to dominate the power output for small $\Delta S$, see fig. \ref{fig:emp}b, which for sufficiently large $w$ and increasing $\Delta S$ gives rise to two local maxima in $\Pout$ as shown in fig. \ref{fig:emp}c. As $\Delta S$ is increased further, the global maximum of the output power again coincides with the local maximum corresponding to $\theta_1^+$, see fig. \ref{fig:emp}d, and the resulting EMP depicted in fig. \ref{fig:emp}a is thus discontinuous. Finally, as $w$ is increased beyond a critical value, the barrier associated to $\theta_2^+$ dominates the output $\Pout$ even as $\Delta S\to \infty$. As a result, the EMP is continuous and qualitatively equal to that obtained for a system with one transition state positioned at $\theta_2^+$ (plus symbols) as illustrated in fig. \ref{fig:emp}a.

\paragraph{Conclusions.}
Traditional engines most often rely on chemical or thermal energy sources to perform mechanical work. In contrast, we have studied a model engine driven by purely entropic forces arising from the topology of the underlying phase space. Our model applies to the large class of physical systems in which a slow reaction coordinate is associated with an exponentially increasing phase space volume. We have shown that force generation in such entropically fuelled systems is surprisingly robust, and that the efficiency in the regime of maximum power operation exhibits novel features such as discontinuities. The proposed model provides, for example, an appropriate framework for describing the unfolding or refolding of a biopolymer chain under the action of an externally controlled force applied by an AFM cantilever or by optical tweezers with the slow coordinate related to the end-to-end length of the polymer \cite{Nelson2007,Imparato2006,Imparato2008,Collin2005}. The model can also be straightforwardly generalized to systems with several coarse-grained coordinates in which larger entropic driving forces can be generated. In conclusion, our work suggests that the important question of the interplay between the energy landscape and the topology in general network structures could be successfully adressed in the future in the context of stochastic thermodynamics. 

\acknowledgements
NG and AI gratefully acknowledge financial support from Lundbeck Fonden. NG also wishes to thank AUFF (Aarhus University), Niels Bohr Fondet, Augustinusfonden and Oticon Fonden for partial financial support. ME is supported by the National Research Fund, Luxembourg under project FNR/A11/02.

\appendix*
\section{Appendix - Continuous phase space formalism}
In the following we demonstrate using the continuous phase space framework how entropic forces arising from an exponential increase in the phase space volume can generate motion against a mechanical gradient.

We consider an overdamped Brownian motion in a two-dimensional potential energy landscape $V(x,y)$ as, e.g., the one depicted in fig. \ref{fig:steppot}, where $x$ represents the macroscopic degree of freedom corresponding to a reaction coordinate affected by a mechanical load force, and $y$ is the degree of freedom to be coarse-grained. The probability distribution function (PDF) $P(x,y)$ describing diffusion in the potential $V(x,y)$ obeys the Fokker-Planck equation,
\begin{equation}
  \label{eq:FP}
  \D{t} P(x,y)=- \nabla \cdot \vec{J}(x,y),
\end{equation}
where the components of the probability current vector $\vec{J}=(J_x,J_y)$ are given by
\begin{equation}
  \label{eq:J}
  J_i=- \sum_{j=x,y} \left[ D_{ij}(\D{j} V+\D{j})\right] P, \quad i=x,y.
\end{equation}
Here, $D_{ij}$ denote the components of the diffusion matrix \cite{vanKampen}. 

Under the assumption of time scale separation between the two coordinates with $y$ being the fast coordinate, the PDF can be written as
\begin{equation}
  P(x,y)=P(x) \pcal(y|x) \simeq P(x)\peq(y|x),
\end{equation}
where the conditional probability for being at $y$ given $x$ attains the equilibrium value
\begin{equation}
  \label{eq:peq}
  \peq(y|x)=e^{-\beta[V(x,y)-\FE(x)]}.
\end{equation}
Here, we have defined the local free energy at $x$,
\begin{equation}
  \label{eq:F}
  \FE(x)=-\beta^{-1} \log \int \dup y \, e^{-\beta V(x,y)}.
\end{equation}
Inserting \eqref{eq:peq} into \eqref{eq:J} and tracing out the $y$-coordinate in \eqref{eq:FP} yields the coarse-grained Fokker-Planck equation for the $x$-direction,
\begin{equation}
  \label{eq:effFP}
  \D{t}P(x)=\D{x}\left[D_{xx}(\D{x}\FE(x)+\D{x}) \right]P(x),
\end{equation}
where the effective nonequilibrium potential for motion along $x$ is given by the free energy defined in \eqref{eq:F}. In the case of diffusion in a two-dimensional narrow channel, eq. \eqref{eq:effFP} is known as the Fick-Jacobs equation \cite{Jacobs1967,Zwanzig1992,Reguera2006}.

\subsection{Flat potential}
 \begin{figure}
\centering
\includegraphics[width=.8\columnwidth]{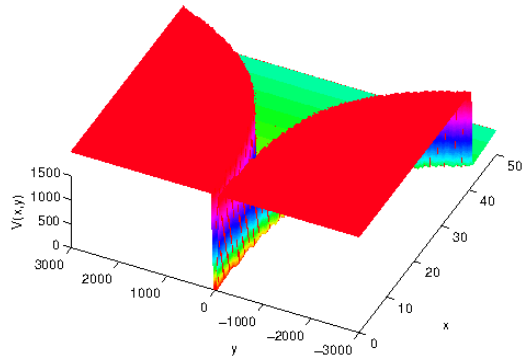}
\caption{A flat potential of the form \eqref{eq:flatpot} with exponentially increasing phase space boundaries $b(x)=b_0\exp[rx]$ for $b_0=50$, $r=0.1$ and $f_0=10$.}
\label{fig:steppot}
\end{figure}
Let us consider as a specific example a potential of the form
\begin{equation}
  \label{eq:flatpot}
  V(x,y)=
\begin{cases} 
U_0(x) \quad \text{if } y\in(-b(x),b(x)) \\
\infty \quad \text{otherwise}
\end{cases}
,
\end{equation}
where the potential energy $U_0(x)=f_0x$ along $x$ is linearly increasing with the load force $f_0>0$, and the potential along $y$ is zero within a bounded region of the phase space given by the function $b(x)$, and infinite otherwise. The free energy \eqref{eq:F} thus becomes
\begin{equation}
  \FE(x)=U_0(x)-\beta^{-1} \log[ 2 b(x)],
\end{equation}
and the resulting effective force $f_x$ in the $x$-direction after coarse-graining is 
\begin{equation}
  f_x=-\partial_x\FE(x)=-f_0+\beta^{-1}b'(x)/b(x).
\end{equation}
In the case of linearly increasing boundaries, i.e. $b(x)=b_0x$, the total force is 
\begin{equation}
  f_x=-f_0+1/(\beta x),
\end{equation}
which is positive for $x<x_0=1/(\beta f_0)$ and negative for $x>x_0$. Hence, in the long-time limit the system reaches thermodynamic equilibrium characterized by the mean position $\langle x \rangle=x_0$. In analogy to the case of the discrete system considered in the Letter, we thus conclude that linear phase space growth is not sufficient to generate steady motion along the macroscopic coordinate. However, if the growth of the boundaries is exponential, $b(x)=b_0 e^{rx}$, as depicted in fig. \ref{fig:steppot}, the resulting force, 
\begin{equation}
  f_x=-f_0+r/\beta \equiv -f_0+f_s,
\end{equation}
is positive for values of $r$ larger than the critical value $r^*=f_0\beta$, thus generating a non-vanishing positive steady-state velocity,  $v_x=f_x/D_{xx}$, along $x$. The resulting motor is thus equivalent to an entropic engine pumping particles uphill against a load force $f_0$ with efficiency $\eta=f_0/f_s=r^*/r$, where $f_s=r/\beta$ is the entropic driving force. 

\subsection{Quadratic potential}
 \begin{figure}
\centering
\includegraphics[width=.8\columnwidth]{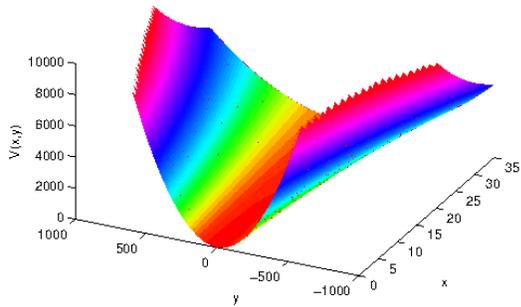}
\caption{Quadratic potential \eqref{eq:quadpot} with $a_0=r=0.05$ and $f_0=10$.}
\label{fig:quadpot}
\end{figure}
Another example of the entropic motor in the continuous phase space formalism is provided by the broadening harmonic potential illustrated in fig. \ref{fig:quadpot} and given mathematically by 
\begin{equation}
\label{eq:quadpot}
  V(x,y)=U_0(x)+a(x)\frac{y^2}{2},
\end{equation}
where the exponential increase of the available phase space in the coarse-grained coordinate is reflected in the function $a(x)=a_0e^{-rx}$. The free energy in this case becomes
\begin{equation}
  \FE(x)=U_0(x)+\frac{1}{2}\beta^{-1}\log\frac{a(x)}{2\pi},
\end{equation}
while the corresponding force 
\begin{equation}
  f_x=-f_0+\beta^{-1}\frac{r}{2}.
\end{equation}
is again able to generate steady-state motion for $r>r^*=2\beta f_0$ with velocity $v_x$ and efficiency $\eta$ as given above.

We have thus shown that application of the coarse-graining approach in stochastic thermodynamics to systems with continuous phase space yields an effective description equivalent to the one obtained on networks. Furthermore, the analysis of the efficiency at maximum power can be carried out in a similar manner in the two classes of systems yielding results qualitatively similar to those in fig. 5 in the main text for an appropriate choice of the potential $V(x,y)$.

%

\end{document}